\documentclass[11pt]{article}%
\usepackage{algorithm}
\usepackage{algpseudocode}
\usepackage{amssymb}
\usepackage{amsfonts}
\usepackage{amsmath}
\usepackage{graphicx}%
\usepackage[caption=false]{subfig}
\providecommand{\U}[1]{\protect\rule{.1in}{.1in}}
\usepackage{tikz}
\usetikzlibrary{shapes}
\usepackage{pgfplots}
\usepackage{verbatim}
\usepgfplotslibrary{statistics}
\pgfplotsset{compat=1.8}
\usepackage{amsmath,amssymb,amsthm}
\usepackage{tikz}
\usetikzlibrary{positioning}

\usepackage{url}

\usepackage{xpatch}
\xpatchcmd\maketitle{\rlap}{}{}{}

\makeatletter
\newcounter{phase}[algorithm]
\newlength{\phaserulewidth}

\newcommand{\phase}[1]{%
  \vspace{-1.25ex}
  \Statex\leavevmode\llap{\rule{\dimexpr\labelwidth+\labelsep}{\phaserulewidth}}\rule{\linewidth}{\phaserulewidth}
  \Statex\strut\refstepcounter{phase}\textit{Step~\thephase~--~#1}
  \vspace{-1.25ex}\Statex\leavevmode\llap{\rule{\dimexpr\labelwidth+\labelsep}{\phaserulewidth}}\rule{\linewidth}{\phaserulewidth}}
\makeatother

\newcommand{\quotes}[1]{``#1''}

\usepackage{graphicx}

\newtheorem{theorem}{Theorem}

\newtheorem{definition}[theorem]{Definition}

\begin{document}

\title{Network-Based Prediction of the 2019-nCoV Epidemic Outbreak in the Chinese Province Hubei}
\author{Bastian Prasse\thanks{Faculty of Electrical Engineering, Mathematics and Computer Science, P.O Box 5031, 2600 GA Delft, The Netherlands; \emph{email}: \{b.prasse, m.a.achterberg, l.ma-2, p.f.a.vanmieghem\}@tudelft.nl}, Massimo A. Achterberg\footnotemark[1], Long Ma\footnotemark[1] and Piet Van Mieghem\footnotemark[1]}
\date{Delft University of Technology \\
 February 12, 2020}
\maketitle

\begin{abstract}
At the moment of writing (12 February, 2020), the future evolution of the 2019-nCoV virus is unclear. Predictions of the further course of the epidemic are decisive to deploy targeted disease control measures. We consider a network-based model to describe the 2019-nCoV epidemic in the Hubei province. The network is composed of the cities in Hubei and their interactions (e.g., traffic flow). However, the precise interactions between cities is unknown and must be inferred from observing the epidemic. We propose a network-based method to predict the future prevalence of the 2019-nCoV virus in every city. Our results indicate that network-based modelling is beneficial for an accurate forecast of the epidemic outbreak. 
\end{abstract}

\section{Introduction}

In December 2019, the novel coronavirus 2019-nCoV emerged in the Chinese city Wuhan \cite{munster2020novel}. Individuals that are infected by the 2019-nCoV virus suffer from the Novel Coronavirus Pneumonia (NCP). Contrary to initial observations \cite{cheng20202019}, the 2019-nCoV virus does spread from person to person as confirmed in \cite{chan2020familial}. On February 12, 2020, there were more than 45,000 confirmed infections, and more than 1000 people died \cite{WHOsituation, ECDCsituation, CDCsituation}. Assessing the further spread of the 2019-nCoV epidemic poses a major public health concern.

Many studies aim to estimate the basic reproduction number $R_0$ of the 2019-nCoV epidemic \cite{zhao2020preliminary,majumder2020early,li2020early,Yang2020.02.10.20021675,imai2019report,liu2020transmission,riou2020pattern,read2020novel,wu2020nowcasting}. The basic reproduction number $R_0$ is a crucial quantity to evaluate the hostility of a virus \cite{hethcote2000mathematics, heesterbeek2002brief}. The basic reproduction number $R_0$ is defined \cite{diekmann1990definition} as \quotes{The expected number of secondary cases produced, in a completely susceptible population, by a typical infective individual during its entire period of infectiousness}. The greater the basic reproduction $R_0$, the more individuals are infected in the long-term endemic state of the virus. If $R_0<1$, then the virus dies out. The estimates for the basic reproduction number $R_0$ of the 2019-nCoV epidemic range from $R_0=2.0$ to $R_0 = 3.77$.

The basic reproduction number $R_0$ only coarsely assesses the quantitative behaviour of the epidemic. To obtain a more detailed picture of the epidemic, the development of epidemic outbreak prediction methods is focal. A diverse body of research considers the prediction of general epidemics. For instance, prediction methods are based on Kalman filtering \cite{yang2014comparison}, Bayesian model averaging \cite{yamana2017individual}, basic regression \cite{brooks2015epiforecast} and kernel density estimation \cite{ray2018prediction}. Recent work focussed on the dependency of population flow and the viral spread \cite{colizza2006role,balcan2009multiscale,belik2011natural,brockmann2013hidden}. As shown by Pei \textit{et al.} \cite{pei2018forecasting}, the spread of influenza can be more accurately predicted by taking the population flow between cities into account. Read \textit{et al.} \cite{read2020novel} predicted the 2019-nCoV epidemic by using the Official Aviation Guide (OAG) Traffic Analyser dataset. Additionally to the OAG dataset, Wu \textit{et al.} \cite{wu2020nowcasting} used the Tencent database to predict the 2019-nCoV viral spread. 

The population flow clearly has an impact on the evolution of an epidemic. However, the exact population flow is unknown, and epidemic prediction methods must account for inaccuracies of population flow data. In this work, we consider the most extreme case by assuming no prior knowledge of the population flow. To forecast the 2019-nCoV epidemic, we design a network-based prediction method that estimates the interactions between cities as an intermediate step. On February 9th, 2020, approximately $70\%$ of the global 2019-nCoV infections are located in the Chinese province Hubei. Thus, we focus on the 2019-nCoV epidemic in Hubei. Our goal is to predict the 2019-nCoV outbreak for every city in Hubei. Section~\ref{sec:data} introduces the available data on the 2019-nCoV virus in Hubei. The epidemic model is proposed in Section~\ref{sec:model}, and the prediction method is outlined in Section~\ref{sec:prediction}. The prediction accuracy is evaluated on past data in Section~\ref{sec:evaluation}. 

\section{Data on the 2019-nCoV Epidemic Outbreak in Hubei}\label{sec:data}
\begin{figure}[!ht]
	\centering
	\includegraphics[width=0.9\textwidth]{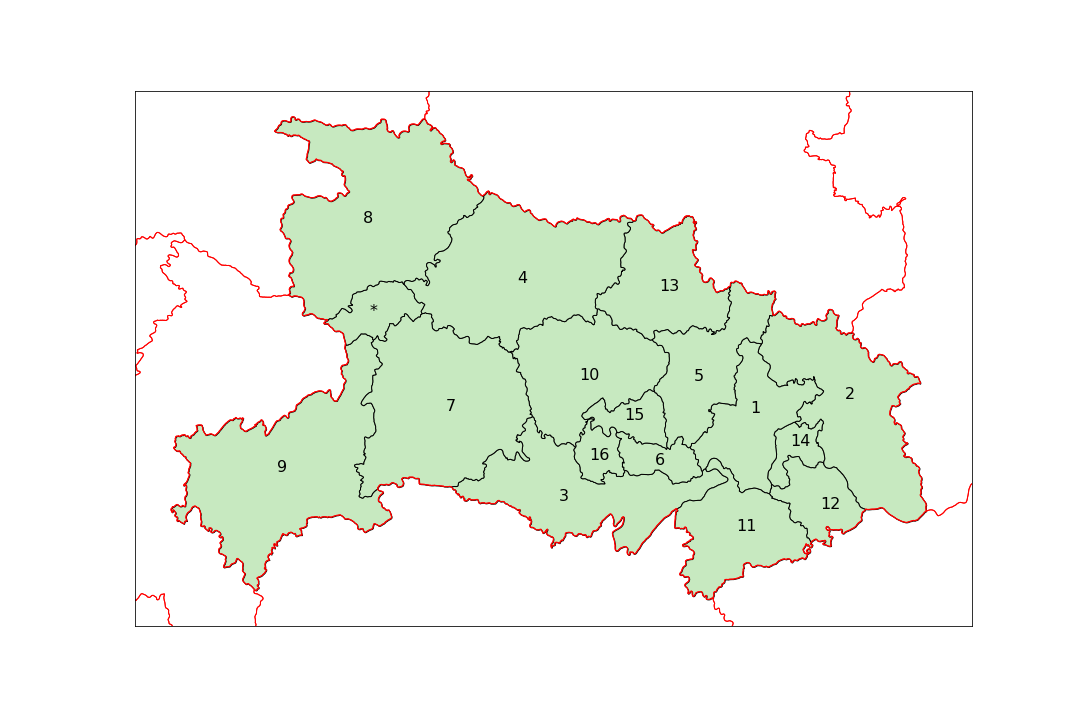}
	\caption{The 17 cities (prefecture-level divisions) of the Chinese province Hubei. The names of the cities are stated in Appendix~\ref{appendix:data_details}. We do not consider the city Shennongjia in this work, which is marked with a star (*).}
	\label{figure:map_hubei}
\end{figure}

The time series of reported infections in Hubei forms the basis for the epidemic outbreak prediction. Hubei is divided into 17 cities (more precisely, prefecture-level divisions) and contains the city Wuhan, as illustrated by Figure~\ref{figure:map_hubei}. We do not consider the city Shennongjia, since the number of infections in Shennongjia is small. We denote the number of considered cities by $N=16$. The number of newly reported infections for each city in Hubei is openly accessible via the website of the Hubei Province Health Committee \cite{hubeiGovernment}. The data is updated daily and follows the standard time offset of UTC+08:00. Except for Wuhan, the total number of reported infections is small before January 21, 2020. Hence, we consider the 2019-nCoV epidemic outbreak starting from January 21. We denote the discrete time by $k\in \mathbb{N}$. The difference of time $k$ to $k+1$ equals one day, and the initial time $k=1$ corresponds to January 21, 2020. The website \cite{hubeiGovernment} states the number of reported infections $N_{\textrm{rep}, i}[k]$ at every time $k$ in every city $i=1, ..., N$. We obtain the population size $p_i$ of each city $i$ from the Hubei Statistical Yearbook \cite{kai2016hubeistayear}. The reported fraction of infected individuals in city $i$ at time $k$ follows as $\mathcal{I}_{\textrm{rep}, i}[k] = N_{\textrm{rep}, i}[k]/p_i$. Appendix~\ref{appendix:data_details} states the population size $p_i$ and the complete time series of the number of infections $N_{\textrm{rep}, i}[k]$ for each city in Hubei. 

\section{Modelling the 2019-nCoV Epidemic between Cities} \label{sec:model}
We model the spread of the 2019-nCoV virus by the SIR-model: At any discrete time $k$, every individual is in either one of the compartments \textit{susceptible} (healthy), \textit{infectious} or \textit{removed}. Susceptible individuals can get infectious due to contact with infectious individuals. Due to hospitalisation, quarantine measures or death, infectious individuals become removed individuals, which cannot infect susceptible individuals any longer. For every city $i$, we denote the $3\times 1$ \textit{viral state} vector at time $k$ by $v_i[k] = (\mathcal{S}_i[k], \mathcal{I}_i[k], \mathcal{R}_i[k] )^T$. The components $\mathcal{S}_i[k]$, $\mathcal{I}_i[k]$, and $\mathcal{R}_i[k]$ denote the fraction of susceptible, infectious, and removed individuals, respectively. Thus, it holds that $\mathcal{S}_i[k]+\mathcal{I}_i[k]+\mathcal{R}_i[k] = 1$ for every city $i$ at every time $k$. The discrete-time SIR model\footnote{The discrete-time SIR epidemic model (\ref{SIR_MF_discrete}) follows from applying Euler's method to the continuous-time mean-field SIR model of Youssef and Scoglio \cite{youssef2011individual}.} is defined as follows.

\begin{definition}[SIR Epidemic Model \cite{youssef2011individual, prasse2019gemf}] 
The viral state $v_i[k] = (\mathcal{S}_i[k], \mathcal{I}_i[k], \mathcal{R}_i[k] )^T$ of every city~$i$ evolves in discrete time $k=1, 2, ...$ according to
	\begin{align} \label{SIR_MF_discrete}
		\mathcal{I}_i[ k + 1] &= (1 - \delta_i) \mathcal{I}_i[k] + (1 - \mathcal{I}_i[k] - \mathcal{R}_i[k]) \sum^N_{j=1} \beta_{ij} \mathcal{I}_j[k] \\  
		\mathcal{R}_i[ k + 1] &= \mathcal{R}_i[k] +  \delta_i \mathcal{I}_i[k], \nonumber
	\end{align}
	and the fraction of susceptible individuals follows as 
	\begin{align}
		\mathcal{S}_i[k] = 1 -\mathcal{I}_i[k] - \mathcal{R}_i[k].\nonumber
	\end{align}
	Here, $\beta_{ij}$ denotes the \emph{infection probability} from city $j$ to city $i$, and $\delta_i$ denotes the \emph{curing probability} of city $i$.
\end{definition}

The SIR model (\ref{SIR_MF_discrete}) assumes that the spreading parameters $\delta_i$, $\beta_{ij}$ do not change over time $k$. The curing probability $\delta_i$ quantifies the capacity of individuals in city $i$ to cure from the virus. The infection probability $\beta_{i j}$ specifies the number of contacts of individuals in city $j$ with individuals in city $i$. We emphasise that $\beta_{ii}\neq 0$ since individuals within one city $i$ do interact with each other. The \textit{contact network} between cities in Hubei is given by the $N\times N$ matrix 
\begin{align}\nonumber
B=\begin{pmatrix}
\beta_{11} & \beta_{12}& ... & \beta_{1N}\\
\vdots & \vdots & \ddots & \vdots\\
\beta_{N1} & \beta_{N2}& ... & \beta_{NN}
\end{pmatrix},
\end{align}
whose elements are probabilities $0 \le \beta_{ij}\le 1$. Neither the curing probabilities $\delta_i$ nor the infection probabilities $\beta_{i j}$ are known for the 2019-nCoV epidemic. Potentially, it is possible to state bounds or estimates for the spreading parameters $\delta_i$ and $\beta_{ij}$ by making use of the people flow or geographical distances between the respective cities. Nevertheless, there would remain an uncertainty regarding the precise value of the spreading parameters $\delta_i$ and $\beta_{ij}$. In this work, we consider the most extreme case: there is no a priori knowledge on the curing probabilities $\delta_i$ nor the infection probabilities $\beta_{ij}$. In Section~\ref{sec:prediction}, we develop an inference method to estimate the spreading parameters $\delta_i$ and $\beta_{ij}$ from observing the epidemic.

\section{Network-Based Approach for Epidemic Outbreak Prediction}\label{sec:prediction}
We propose a network-based method to predict the outbreak of 2019-nCoV virus, which consists of three steps. First, we preprocess the raw data of the confirmed number of infected individuals in Subsection~\ref{subsec:data_preprocess} to obtain an SIR time series $v_i[1], v_i[2], ...$ of the viral state for every city $i$. Second, based on the time series $v_i[1], v_i[2], ...$, we obtain estimates $\hat{\delta}_i$ and $\hat{\beta}_{ij}$ of the unknown spreading spreading parameters $\delta_i$ and $\beta_{ij}$ in Subsection~\ref{subsec:network_reconstruction}. Third, the estimates $\hat{\delta}_i$ and $\hat{\beta}_{ij}$ result in an SIR model (\ref{SIR_MF_discrete}), which we iterate for future times $k$ to predict the evolution of the 2019-Cov virus. Subsection~\ref{subsec:data_preprocess} and Subsection~\ref{subsec:network_reconstruction} give an outline of the first two steps of the prediction method. We refer the reader to Appendix~\ref{appendix:prediction_algorithm} for a detailed description of the prediction method.

\subsection{Data Preprocessing}\label{subsec:data_preprocess}
We denote the number of observations by $n$, which equals the number of days since January 21, 2020. Our goal is to obtain an SIR viral state vector $v_i[k]= (\mathcal{S}_i[k], \mathcal{I}_i[k], \mathcal{R}_i[k] )^T$ for every city $i$ at any time $k=1, ..., n$ based on the data described in Section~\ref{sec:data}. The fraction of susceptible individuals follows as $\mathcal{S}_i[k] = 1 - \mathcal{I}_i[k] - \mathcal{R}_i[k]$ at any time $k\ge 1$. Thus, it suffices to determine the fraction of infectious individuals $\mathcal{I}_i[k]$ and recovered individuals $\mathcal{R}_i[k]$. The fraction of infectious individuals $\mathcal{I}_i[k]$ follows\footnote{The measurement data in Section~\ref{sec:data} is the number $N_{\textrm{rep}, i}[k]$ of individuals that are \textit{detected} to be infected by 2019-nCoV. Upon detection of the infection, the respective individuals are hospitalised and, hence, not infectious any more to individuals outside of the hospital. We consider the reported fraction of infections $\mathcal{I}_{\textrm{rep}, i}[k]$ as an \textit{approximation} for the number of infectious individuals $\mathcal{I}_i[k]$.} from the reported fraction of infections $\mathcal{I}_{\textrm{rep}, i}[k]$ described in Section~\ref{sec:data}. We emphasise that the reported fraction of infections $\mathcal{I}_{\textrm{rep}, i}[k]$ only lower-bounds the true fraction of infected individuals~$\mathcal{I}_i[k]$ for two reasons. First, not all infectious individuals are aware that they are infected. Second, the diagnosing capacities in the hospitals are limited, particularly when the number of infections increases rapidly. Hence, not all infectious individuals that arrive at a hospital can be reported timely.

We do not know the fraction of removed individuals $\mathcal{R}_i[k]$. At the initial time $k=1$, it is realistic to assume that $\mathcal{R}_i[1]=0$ holds for every city $i$. At any time $k\ge 2$, the removed individuals $\mathcal{R}_i[k]$ could be obtained from (\ref{SIR_MF_discrete}), if the curing probability $\delta_i$ were known. However, we do not know the curing probability $\delta_i$. Hence, we consider 50 equidistant candidate values for the curing probability $\delta_i$, ranging from $\delta_\textrm{min}=0.01$ to $\delta_\textrm{max}=1$. We define the set of candidate values as $\Omega = \{\delta_\textrm{min}, ..., \delta_\textrm{max}\}$. For every candidate value $\delta_i \in \Omega$, the fraction of removed individuals $\mathcal{R}_i[k]$ follows from (\ref{SIR_MF_discrete}) at all times $k\ge 2$. Thus, we obtain 50 potential sequences $\mathcal{R}_i[1], ...,\mathcal{R}_i[n]$, each of which corresponding to one candidate value $\delta_i \in \Omega$. We estimate the curing probability $\delta_i$, and hence implicitly the sequence $\mathcal{R}_i[1], ...,\mathcal{R}_i[n]$, as the element in $\Omega$ that resulted in the best fit of the SIR model (\ref{SIR_MF_discrete}) to the measured number of infections.

The raw time series $\mathcal{I}_{\textrm{rep},i}[1], ..., \mathcal{I}_{\textrm{rep},i}[n]$ exhibits erratic fluctuations. There is a single outlier\footnote{Potentially, the outlier is due to the increase in the maximum number of individuals that can be diagnosed in Wuhan, from 200 to 2000 individuals per day as of January 27th \cite{outlier_chinanews}.} in city $i=1$ (Wuhan) at time $k=17$ (January 28, 2020), which we replace by $\mathcal{I}_{\textrm{rep},1}[17]= (\mathcal{I}_{\textrm{rep},1}[16]+\mathcal{I}_{\textrm{rep},1}[18])/2$. To reduce the fluctuations, we apply a moving average, provided by the Matlab command \texttt{smoothdata}, to the time series $\mathcal{I}_{\textrm{rep},i}[1], ..., \mathcal{I}_{\textrm{rep},i}[n]$ of every city $i$. The preprocessed time series $\mathcal{I}_i[1], ..., \mathcal{I}_i[n]$ equals the output of \texttt{smoothdata}.

\subsection{Network Inference} \label{subsec:network_reconstruction}
For every city $i$, the curing probability $\delta_i$ is estimated as one of the candidate values in $\Omega$, as outlined in Subsection~\ref{subsec:data_preprocess}. The remaining task is to estimate the infection probabilities $\beta_{ij}$. The goal of \textit{network inference} \cite{ma2019inferring,di2019network, timme2014revealing, wang2016data} is to estimate the matrix of infection probabilities $B$ from the SIR viral state observations $v_i[1], ..., v_i[n]$. The matrix $B$ can be interpreted as a weighted adjacency matrix. We adapt a network inference approach\footnote{The network inference approach \cite{prasse2019gemf} is also applicable to general compartmental epidemic models \cite{sahneh2013generalized}, such as the Susceptible-Exposed-Infected-Removed (SEIR)
epidemic model.} \cite{prasse2018networkreconstruction,prasse2019gemf}, which is based on formulating a set of linear equations and the \textit{least absolute shrinkage and selection operator} (LASSO) \cite{tibshirani1996regression, hastie2015statistical}. The crucial observation from the SIR governing equations (\ref{SIR_MF_discrete}) is that $\beta_{ij}$ appears linearly, whereas the state variables $\mathcal{S}_i$, $\mathcal{I}_i$ and $\mathcal{R}_i$ do not. From (\ref{SIR_MF_discrete}), the infection probabilities $\beta_{ij}$ satisfy 
\begin{align}\label{linear_system}
V_i = F_i  \begin{pmatrix}
\beta_{i1} \\
\vdots \\
 \beta_{iN}
\end{pmatrix}
\end{align}
for all cities $i=1, ..., N$. Here, the $(n-1) \times 1$ vector $V_i$ and the $(n-1) \times N$ matrix $F_i$ are given by
\begin{align}\label{V_i}
V_i = \begin{pmatrix}
\mathcal{I}_i[2] - (1 - \delta_i)\mathcal{I}_i[1]\\
\vdots \\
\mathcal{I}_i[n] - (1 - \delta_i)\mathcal{I}_i[n-1]
\end{pmatrix}
\end{align}
and
\begin{align}\label{F_i}
F_i = \begin{pmatrix}
\mathcal{S}_i[1] \mathcal{I}_1[1]& ... & \mathcal{S}_i[1] \mathcal{I}_N[1] \\
\vdots & \ddots & \vdots\\
\mathcal{S}_i[n-1] \mathcal{I}_1[n-1]& ... & \mathcal{S}_i[n-1] \mathcal{I}_N[n-1] 
\end{pmatrix}.
\end{align}
If the SIR model (\ref{SIR_MF_discrete}) were an exact description of the evolution of the coronavirus, then the linear system (\ref{linear_system}) would hold with equality. However, the viral state vector $v_i[k]$ in city $i$ does not exactly follow the SIR model (\ref{SIR_MF_discrete}). Instead, the evolution of the viral state vector $v_i[k]$ is described by 
\begin{align}\nonumber
v_i [k + 1] & = f_\textrm{SIR}(v_1[k], ..., v_N[k]) + w_i[k],
\end{align}
where the $3\times 1$ vector $f_\textrm{SIR}(v_1[k], ..., v_N[k])$ denotes the right-hand sides of the SIR model (\ref{SIR_MF_discrete}), and the $3\times 1$ vector $w_i[k]$ denotes the unknown \textit{model error} of city $i$ at time $k$. Due to the model errors~$w_i[k]$, the linear system (\ref{linear_system}) only holds approximately. Thus, we resort to estimating the infection probabilities $\beta_{ij}$ by minimising the deviation of the left side and the right side of (\ref{linear_system}). We reconstruct the network by the LASSO \cite{tibshirani1996regression, hastie2015statistical} as follows:
\begin{align}\label{lasso}
\begin{aligned} 
 & \underset{\beta_{i1}, ..., \beta_{iN}}{\operatorname{min}}  & & \left\lVert V_i - F_i \begin{pmatrix}
\beta_{i1} \\
\vdots \\
 \beta_{iN}
\end{pmatrix} \right\rVert^2_2 + \rho_i \sum^N_{j=1, j\neq i}\beta_{ij} & \\
 &\text{s.t.} & & 0\le \beta_{ij} \le 1, \quad j=1, ..., N. &
\end{aligned} 
 \end{align}
 The first term in the objective function of (\ref{lasso}) measures the deviation of the left side and the right side of (\ref{linear_system}). The sum in the objective of (\ref{lasso}) is an $\ell_1$--norm regularisation term which avoids overfitting. We choose to not penalise the self-infection probability $\beta_{ii}$, since we expect the infections among individuals within the same city $i$ to be dominant. The regularisation parameter $\rho_i>0$ is set by cross--validation. The LASSO network inference (\ref{lasso}) allows for the incorporation of a priori knowledge of the contact network $B$ by adding further constraints to the infection probabilities $\beta_{ij}$. We emphasise that an accurate prediction of an SIR epidemic outbreak does not require an accurate network inference~\cite{prasse2019gemf}. 

\section{Evaluation of the Prediction Accuracy} \label{sec:evaluation}
The accuracy of the network-based prediction method in Section~\ref{sec:prediction} is evaluated by comparison to a simple prediction method. Qualitatively, the virus spread in many epidemiological model follows a sigmoid function, see also \cite{van2016universality}. A particular sigmoid function is obtained by logistic regression. As a comparison to the method in Section~\ref{sec:prediction}, we apply logistic regression on the reported fractions $\mathcal{I}_{\textrm{rep}, i}[1]$, ..., $\mathcal{I}_{\textrm{rep}, i}[n]$ of infection individuals, \emph{independently} for each city $i$ in Hubei. Logistic regression is advantageous because a logistic function is a closed-form expression, and its parameters can be determined by non-linear regression. Moreover, the logistic function is an approximation to the exact solution of some epidemiological models and population growth models \cite{verhulst1838notice, van2016universality,prasse2019time}. For further details regarding logistic regression, we refer the reader to Appendix~\ref{appendix:logistic_regression}.

We denote the cumulative fraction of infections at time $k$ by
\begin{equation}\nonumber
\mathcal{I}_{\textrm{cs}, i} [k] = \sum^k_{\tau=1} \mathcal{I}_i [\tau].
\end{equation}

\begin{figure*}[!ht]
	\centering
	\subfloat[$m=1$ day\label{figure:estimate1dayprior}]{%
		\includegraphics[width=0.49\textwidth]{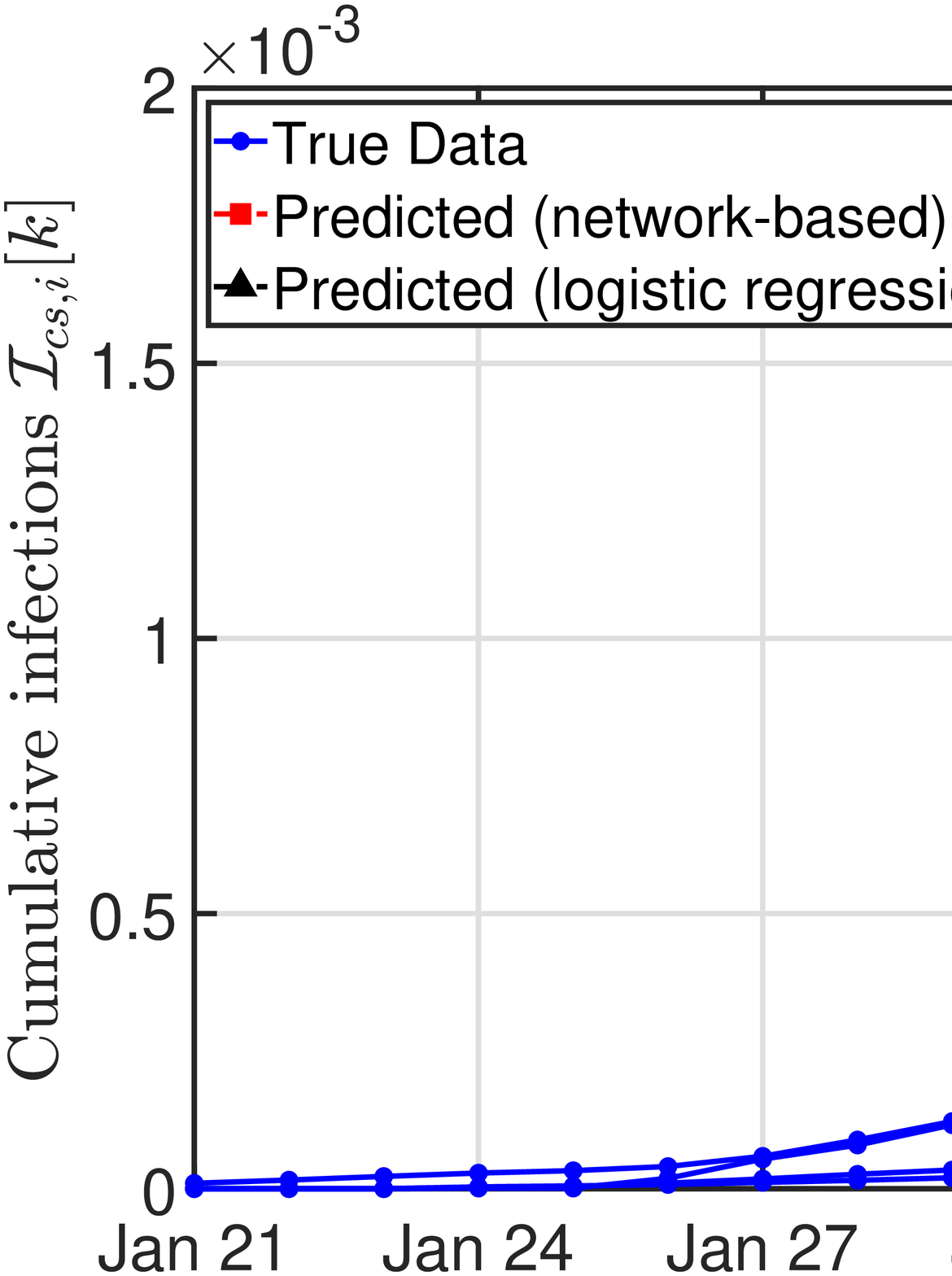}
	}
	\subfloat[$m=2$ days\label{figure:estimate2dayprior}]{%
		\includegraphics[width=0.49\textwidth]{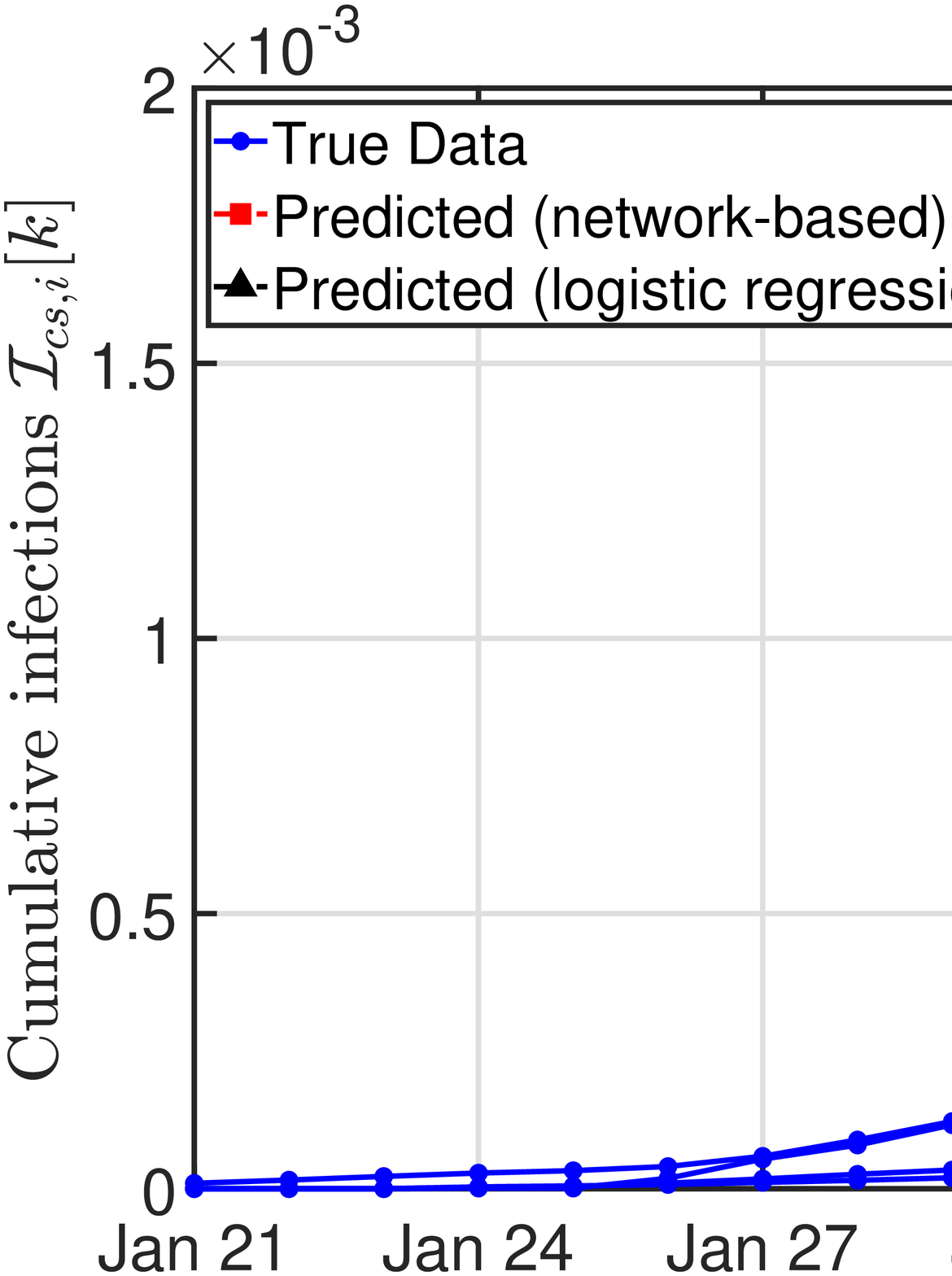}
	}\\
	\subfloat[$m=3$ days\label{figure:estimate3dayprior}]{%
		\includegraphics[width=0.49\textwidth]{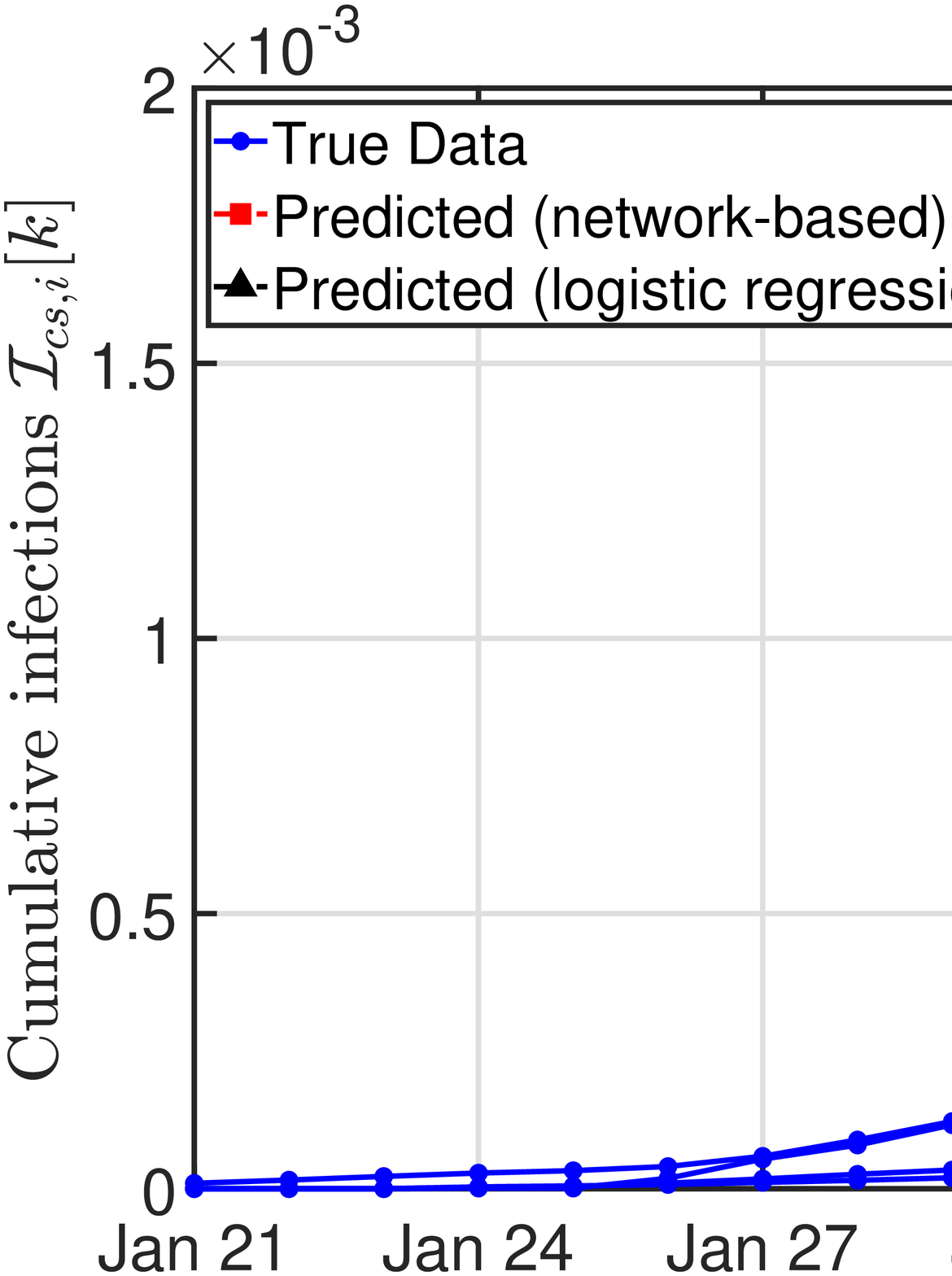}
	}
	\subfloat[$m=4$ days\label{figure:estimate4dayprior}]{%
		\includegraphics[width=0.49\textwidth]{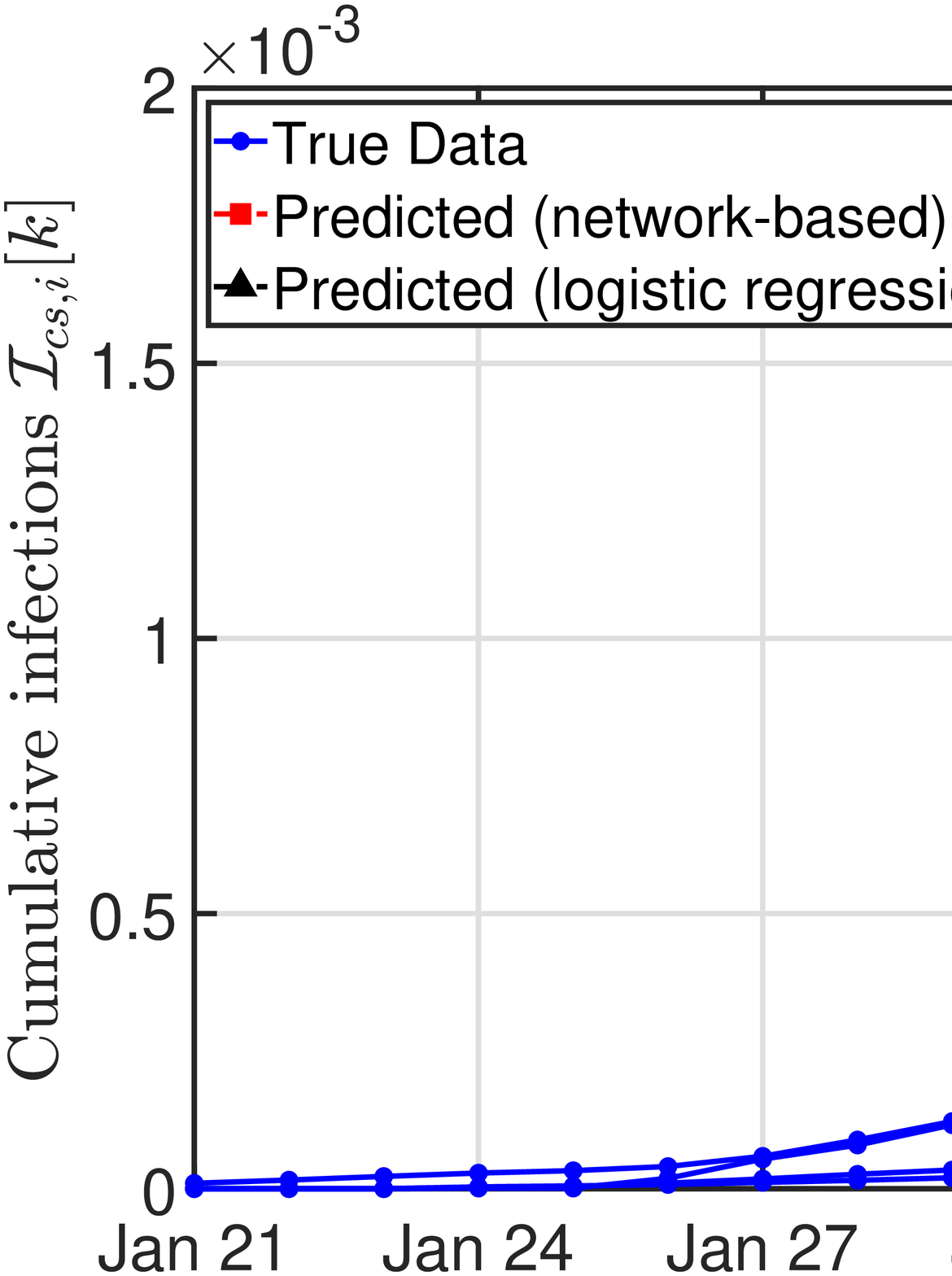}
	}
	\caption{The prediction of the 2019-nCoV outbreak in Hubei by the network-based prediction method~(\ref{lasso}) and by simple logistic regression. For clarity, only four of the $N=16$ cities are depicted. Each subfigure is obtained by omitting a number $m=1, 2, 3, 4$ of days prior to February 11, 2020, and subsequently predicting the same number of days ahead in time. The omitted number of data points is equal to: (a) $m=1$ day, (b) $m=2$ days, (c) $m=3$ days and (d) $m=4$ days. The first prediction data point, for instance February 10 in subfigure (a), coincides with the last day that has been observed.}
	\label{figure:estimateDaysPrior}
\end{figure*}

At the time of writing, the data is available from January 21 until February 11, 2020. To evaluate the prediction accuracy, we remove the data for a fixed number of days, say $m$, prior to February~11. The prediction model is determined upon the data from 21 January up to $11-m$ February, 2020. Then, we predict the course of the disease up to February 11, and the number of omitted days $m$ is equal to the number of prediction days. The course of the disease is shown in Figure \ref{figure:estimateDaysPrior} for the removal of: (a) $m=1$ day, (b) $m=2$ days, (c) $m=3$ days and (d) $m=4$ days.

For most predictions shown in Figure \ref{figure:estimateDaysPrior}, the logistic curve appears to underestimate the true fraction of infected individuals, whereas the network-based method seems to overestimate the true value. The logistic curve is therefore a lower bound prediction for the real fraction of infected individuals. 

\begin{figure*}[!ht]
	\centering
	\subfloat[$m=3$ days\label{figure:estimate1daypriorerror}]{%
		\includegraphics[width=0.49\textwidth]{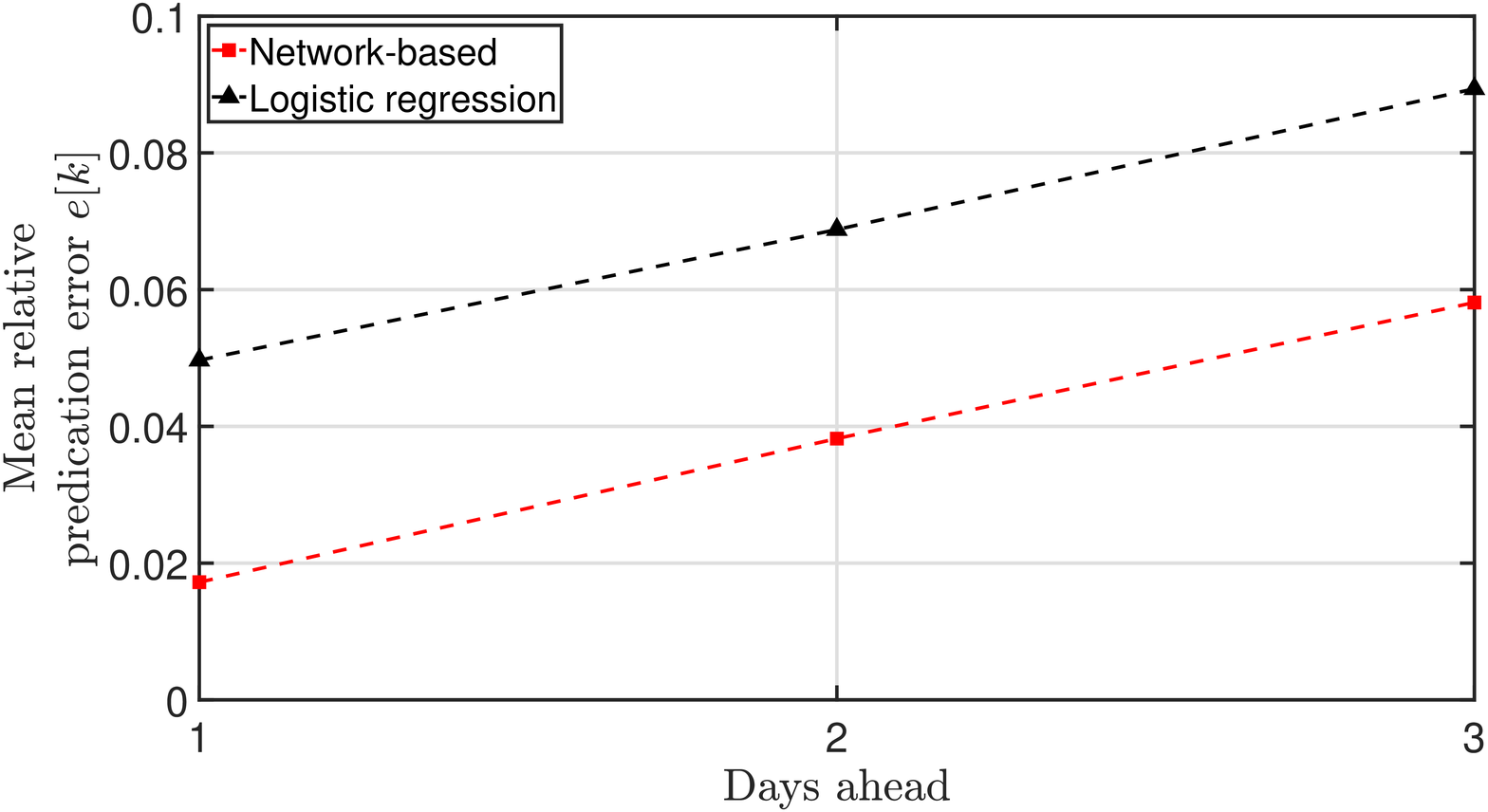}
	}
	\subfloat[$m=4$ days\label{figure:estimate2daypriorerror}]{%
		\includegraphics[width=0.49\textwidth]{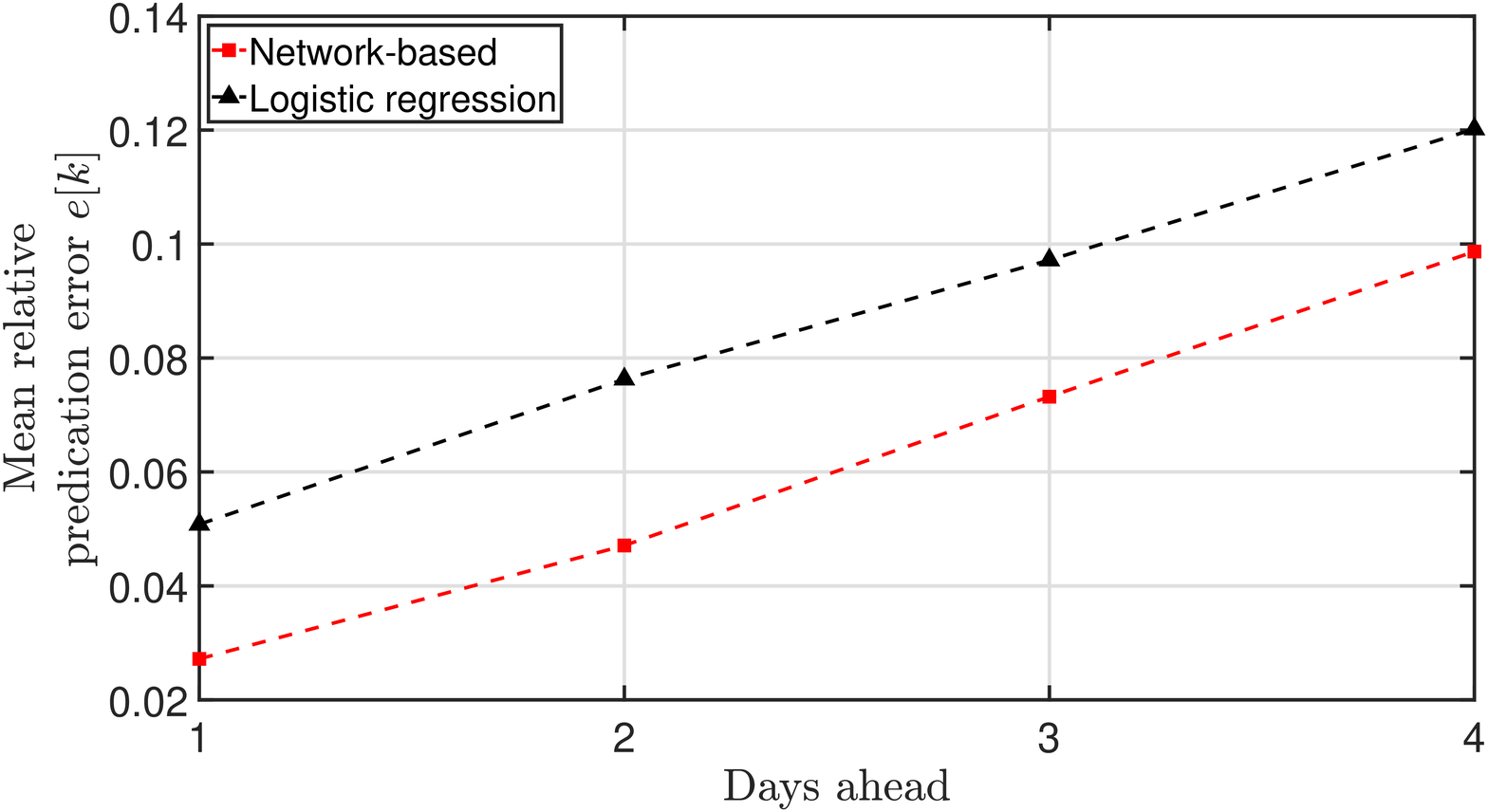}
	}
	\caption{The accuracy of both prediction methods to forecast the 2019-nCoV outbreak in Hubei. Each subfigure is obtained by omitting a number of days prior to February 11, and subsequently predicting the same number of days ahead in time. The removed data points equal to $m=3$ days (a) and $m=4$  days (b).}
	\label{figure:estimateDaysPriorError}
\end{figure*}

The prediction accuracy decreases if the prediction time is increased, which we quantify by the Mean Absolute Percentage Error (MAPE) 
\begin{equation}\nonumber
	e[k] = \frac{1}{N} \sum_{i=1}^N \frac{|\hat{\mathcal{I}}_{\textrm{cs}, i}[k] - \mathcal{I}_{\textrm{cs}, i} [k]|}{\mathcal{I}_{\textrm{cs}, i} [k]}, \label{eq:errormetric}
\end{equation}
at any time $k$. Here, $\hat{\mathcal{I}}_{\textrm{cs}, i}[k]$ denotes the predicted cumulative fraction of individuals of city $i$ at time~$k$. Figure \ref{figure:estimateDaysPriorError} depicts the MAPE prediction error for the data shown in Figure \ref{figure:estimateDaysPrior}. Two observations are worth mentioning. First, as expected, the prediction error increases when predicting more days ahead. Second, the network-based method always provides more accurate predictions than the logistic regression.

\begin{figure*}[!ht]
	\centering
	\includegraphics[width=0.95\textwidth]{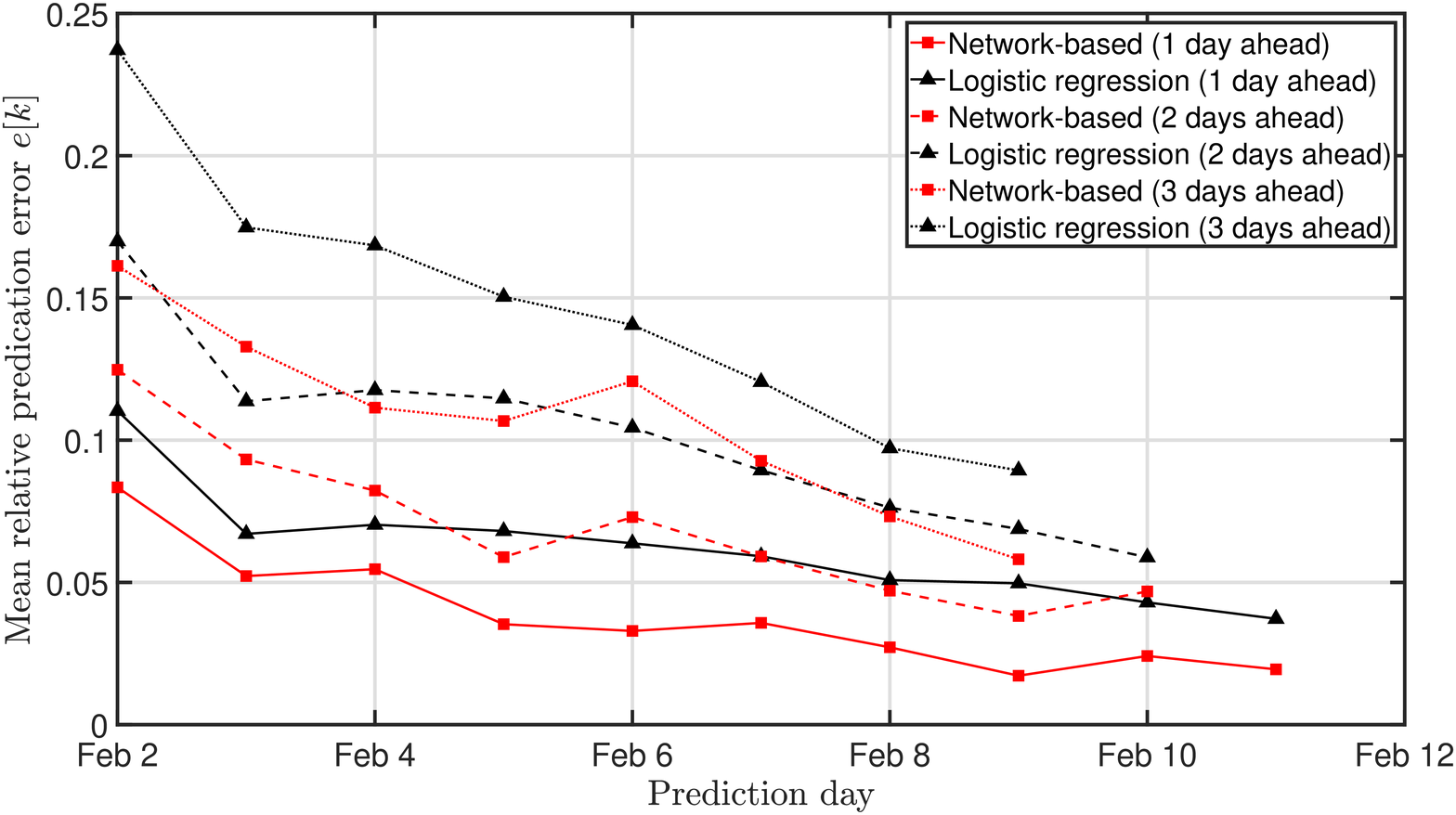}
	\caption{The accuracy of both prediction methods for the 2019-nCoV outbreak versus the date until the data is available. The solid lines correspond to a 1-day ahead prediction of the fraction of infected individuals. Dashed lines correspond to a 2-days ahead prediction, and the dotted lines corresponds to 3-days ahead.}
	\label{figure:estimatePredictionError}
\end{figure*}

Figure \ref{figure:estimatePredictionError} illustrates the prediction accuracy versus the time that the epidemic outbreak has been observed. As the epidemic evolves, the prediction accuracy increases.

Finally, we consider the prediction of the fraction of infected individuals for the next five days. We stress that, as shown in Figure \ref{figure:estimateDaysPrior} and Figure \ref{figure:estimatePredictionError}, the prediction might be inaccurate for more than four days ahead. The predicted number of infected individuals for each city $i$ is shown in Table~\ref{table:outbreakprediction}.

\begin{figure*}[!ht]
	\centering
	\includegraphics[width=0.95\textwidth]{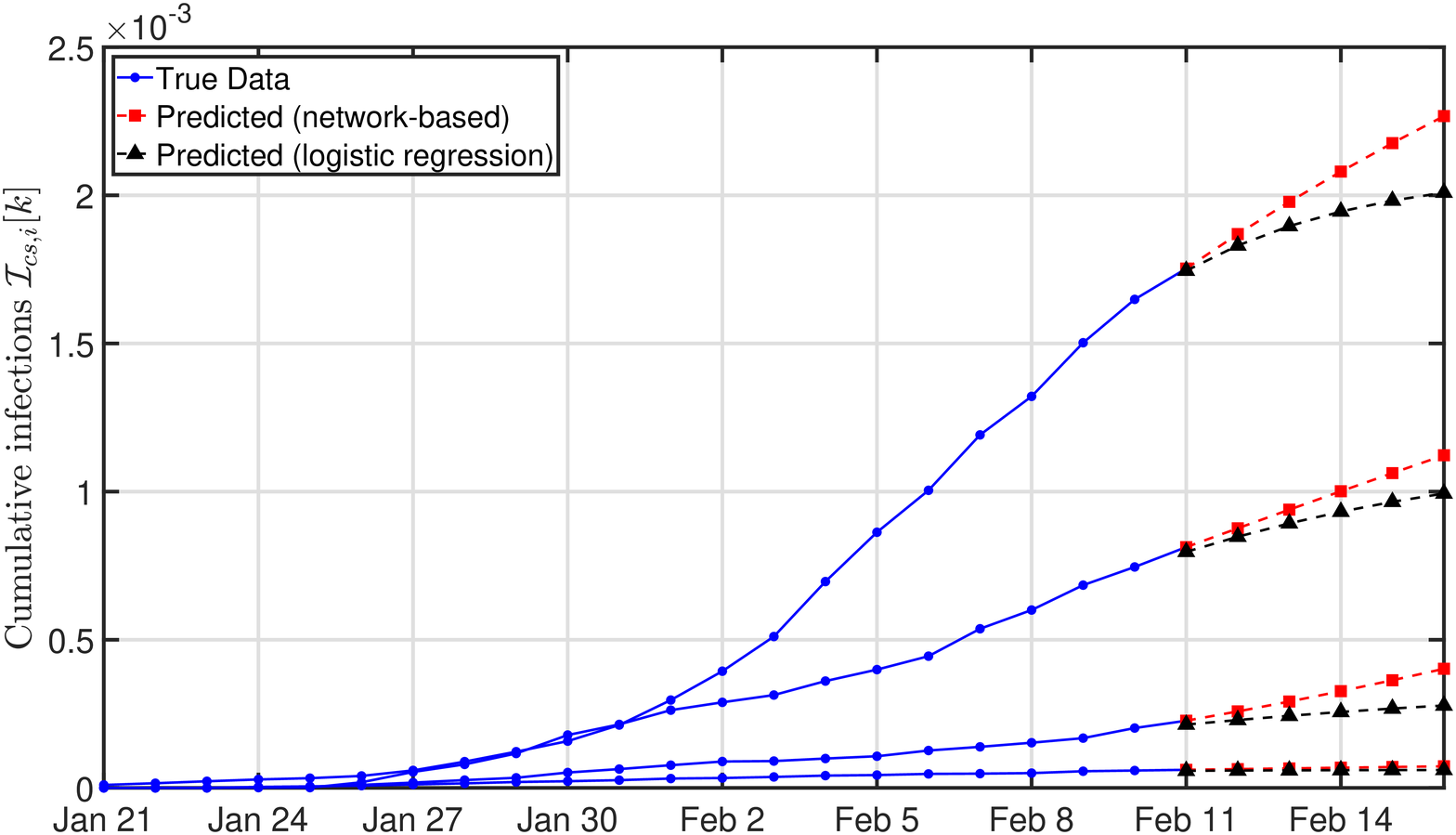}
	\caption{The predicted number of infected individuals for 4 cities in Hubei by both methods for the coming 5 days. The first prediction data point (February 11, 2020) coincides with the last day that has been observed. We selected 4 of the $N=16$ cities to avoid overlapping curves. The predicted number of infected individuals for the network inference method is presented in Table~\ref{table:outbreakprediction}.}
	\label{figure:estimateFuture}
\end{figure*}

\begin{table*}[!ht]
	\centering
	\begin{tabular}{|l|r|r|r|r|r|r|}
		\hline       
		Identifier $i$& City & Feb 12 & Feb 13 & Feb 14 & Feb 15 & Feb 16 \\ \hline \hline 
		1 & Wuhan 		& $19826$ & $20981$ & $22064$ & $23084$ & $24052$ \\ \hline 
		2 & Huanggang 	& $2465$ & $2530$ & $2596$ & $2662$ & $2728$ \\ \hline 
		3 & Jingzhou 	& $1145$ & $1177$ & $1209$ & $1240$ & $1271$ \\ \hline 
		4 & Xiangyang 	& $1119$ & $1149$ & $1179$ & $1209$ & $1239$ \\ \hline 
		5 & Xiaogan 	& $2855$ & $2954$ & $3049$ & $3142$ & $3232$ \\ \hline 
		6 & Xiantao 	& $483$ & $507$ & $530$ & $554$ & $577$ \\ \hline 
		7 & Yichang 	& $804$ & $825$ & $847$ & $870$ & $894$ \\ \hline 
		8 & Shiyan 		& $560$ & $581$ & $601$ & $620$ & $639$ \\ \hline 
		9 & Enshi 		& $211$ & $219$ & $227$ & $235$ & $242$ \\ \hline 
		10 & Jingmen 	& $717$ & $736$ & $755$ & $774$ & $792$ \\ \hline 
		11 & Xianning 	& $534$ & $543$ & $552$ & $561$ & $570$ \\ \hline 
		12 & Huangshi 	& $906$ & $935$ & $962$ & $988$ & $1014$ \\ \hline 
		13 & Suizhou 	& $1164$ & $1198$ & $1230$ & $1262$ & $1295$ \\ \hline 
		14 & Ezhou 		& $928$ & $995$ & $1061$ & $1126$ & $1189$ \\ \hline 
		15 & Tianmen 	& $334$ & $376$ & $422$ & $469$ & $519$ \\ \hline 
		16 & Qianjiang 	& $93$ & $96$ & $98$ & $101$ & $104$ \\ \hline 
	\end{tabular}
	\caption{The cumulative number of infected individuals predicted by the network-based method for each city in Hubei. \label{table:outbreakprediction}} 
\end{table*}

\section{Conclusions}
We applied a network-based SIR epidemic model to predict the outbreak of the 2019-nCoV virus for each city in the Chinese province Hubei. The epidemic model allows to explicitly specify the interactions of individuals of different cities, for instance by using traffic patterns between cities. However, the precise interactions between cities is unknown and must be inferred from observing the evolution of the epidemic.

We proposed a network-based prediction method, which estimates the interactions between cities as an intermediate step. We did not assume any prior knowledge on the interactions between cities. The prediction method is evaluated on past data of the 2019-nCoV outbreak in Hubei. Our results indicate that a network-based modelling approach may yield more accurate predictions than modelling the epidemic for each city independently. We believe that the prediction method can be further improved, e.g., by using traffic flow patterns as prior knowledge. 

\section*{Acknowledgements}
We are grateful to Fenghua Wang for helping with collecting the data. Long Ma is grateful for the support from the China Scholarship Council.

\appendix
\section{Details on the Data of the 2019-nCoV Epidemic Outbreak} \label{appendix:data_details}
Table~\ref{table:city_names} shows the cities of the province Hubei and the respective population size $p_i$ for every city $i$. The time series of the reported number of infections $N_{\textrm{rep}, i}[k]$ is stated in Table~\ref{table:time_series}.

\begin{center}
  \begin{table*}
  \centering
  \begin{tabular}{ | l | l | r | }
    \hline       
Identifier $i$ & City & Population $p_i$ \\ \hline \hline 
1 & Wuhan & 10,607,700 \\ \hline 
2 & Huanggang & 6,291,000 \\ \hline 
3 & Jingzhou & 5,705,900 \\ \hline 
4 & Xiangyang & 5,614,000 \\ \hline 
5 & Xiaogan & 4,878,000 \\ \hline 
6 & Xiantao & 1,155,000 \\ \hline 
7 & Yichang & 4,115,000  \\ \hline 
8 & Shiyan& 3,383,000 \\ \hline 
9 & Enshi (autonomous prefecture) & 3,327,000 \\ \hline 
10 & Jingmen& 2,896,300  \\ \hline 
11 & Xianning & 2,507,000\\ \hline 
12 & Huangshi & 2,458,000\\ \hline 
13 & Suizhou & 2,190,800\\ \hline 
14 & Ezhou & 1,059,500\\ \hline 
15 & Tianmen & 1,292,000\\ \hline 
16 & Qianjiang & 958,000\\ \hline 
  \end{tabular}
   \caption{Cities (prefecture-level divisions) in the province Hubei, China. We do not consider the city Shennongjia in this work, since the number of infections with the 2019-nCoV in Shennongjia is very small. \label{table:city_names}} 
  \end{table*}
\end{center}

\begin{table}[p]
	\centering
	\rotatebox{90}{
	\tiny
		\begin{minipage}{\textheight}
\begin{tabular}{|c|c|c|c|c|c|c|c|c|c|c|c|c|c|c|c|c|c|c|c|c|c|c|} 
\hline
City&21-1&22-1&23-1&24-1&25-1&26-1&27-1&28-1&29-1&30-1&31-1&1-2&2-2&3-2&4-2&5-2&6-2&7-2&8-2&9-2&10-2&11-2\\       
\hline
Wuhan&105&62&70&65&46&80&892&315&356&378&576 &894&1033&1242&1967&1766&1501&1985&1378&1921&1552
&1104\\\hline
Huanggang&0	&0	&0	&64	&58	&32	&59	&111	&172	&77	&153	&276	&244	&176	&223	&162	&90	&144	&96	&115&80&66\\\hline 
Jingzhou&0	&6	&2	&2	&23	&14	&24	&30	&50	&70	&66	&46	&166	&114	&100	&88	&84	&56	&56	&48&30&35\\\hline    
Xiangyang&0	&0	&0	&0	&2	&34	&34	&61	&32	&123	&61	&94	&107	&84	&103	&52	&51	&69	&55	&57&44&25\\\hline    
Xiaogan&0	&0	&22	&4	&29	&45	&73	&101	&125	&142	&87	&121	&169	&202	&342	&424	&255	&172	&123	&105&101&109\\\hline    
Xiaotao&0	&0	&2	&8	&1	&1	&15	&5	&23	&35	&7	&43	&29	&19	&37	&40	&42	&52	&20	&37&22&22\\\hline    
Yichang&0	&0	&1	&0	&19	&11	&20	&12	&54	&50	&109	&77	&39	&60	&44	&67	&47	&23	&71	&45&23&12\\\hline    
Shiyan&0	&0	&1	&4	&15	&20	&25	&23	&31	&31	&27	&35	&44	&35	&27	&35	&42	&43	&29	&14&24&31\\\hline    
Enshi&0	&0	&0	&11	&6	&8	&13	&13	&15	&9	&12	&18	&6	&12	&15	&6	&13	&3	&6	&21&8&8\\\hline    
Jingmen&0	&1	&0	&20	&17	&52	&24	&28	&49	&36	&24	&78	&16	&55	&22	&86	&45	&35	&41	&12&15&40\\\hline    
Xianning&0	&0	&0	&0	&43	&21	&27	&21	&18	&36	&40	&40	&50	&52	&36	&15	&44	&33	&17	&14&8&10\\\hline    
Huangshi&0	&0	&0	&0	&31	&5	&17	&33	&27	&55	&41	&43	&82	&71	&104	&57	&69	&68	&50	&52&30&39\\\hline    
Suizhou&0	&0	&0	&5	&31	&16	&18	&46	&27	&85	&76	&80	&74	&183	&65	&128	&81	&38	&31	&65&46&34\\\hline    
Ezhou&0	&0	&0	&1	&0	&19	&37	&27	&39	&66	&38	&51	&28	&26	&50	&41	&48	&98	&67	&89&65&71\\\hline
Tianmen&0	&0	&0	&3	&2	&8	&10	&11	&10	&23	&15	&17	&16	&2	&11	&10	&25	&16	&18	&20&44&32\\\hline    
Shennongjia&0	&0	&0	&0	&0	&0	&1	&2	&2	&2	&0	&0	&0	&3	&0	&0	&0	&0	&0	&0&0&0\\\hline    
Qianjiang&0	&0	&0	&0	&0	&5	&2	&1	&2	&2	&15	&8	&0	&9	&10	&10	&10	&6	&2	&3&5&0\\\hline       
\end{tabular}\caption{The time series of the reported number of infections $N_{\textrm{rep}, i}[k]$ for every city $i$. \label{table:time_series}}
		\end{minipage}
}
\end{table}

\section{Details of the Prediction Method} \label{appendix:prediction_algorithm}

Algorithm~\ref{algorithm:prediction} describes the prediction method, which was outlined in Section~\ref{sec:prediction}, in pseudocode\footnote{The Matlab code is available upon request to the authors.}. In line~4, the Matlab command \texttt{smoothdata} is called to remove erratic fluctuations of the raw data~$\mathcal{I}_{\textrm{rep}, i}[k]$. We denote the $N\times 1$ infection state vector by $\mathcal{I}[k] = (\mathcal{I}_1[k], ..., \mathcal{I}_N[k])^T$ at any time $k$. The loop starting in line~8 iterates over all candidate values of the curing probability $\delta_i$ which are in the set $\Omega$. Algorithm~\ref{algorithm:prediction} calls the network inference method, which is stated in pseudocode by Algorithm~\ref{algorithm:reconstruction}. For a fixed curing probability $\delta_i$, the network inference in line~12 returns an estimate for the infection probabilities $\beta_{i1}\left( \delta_i \right)$, ..., $\beta_{iN}\left( \delta_i \right)$. Furthermore, the network inference returns the mean squared error~$\operatorname{MSE}\left( \delta_i \right)$, which corresponds to the first term in the objective of (\ref{lasso}). The smaller the mean squared error $\operatorname{MSE}\left( \delta_i \right)$, the better the fit of the SIR model (\ref{SIR_MF_discrete}) to the data $\mathcal{I}_i[1], ..., \mathcal{I}_i[n]$. In line~14, the final estimate $\hat{\delta}_i$ for the curing probability is obtained as the minimiser of the mean squared error $\operatorname{MSE}\left( \delta_i \right)$. The estimate $\hat{\delta}_i$ determines the final estimates $\hat{\beta}_{i1}$, ..., $\hat{\beta}_{iN}$ for the infection probabilities in line~15. From line~17 to line~27, the SIR model (\ref{SIR_MF_discrete}) is iterated, which results in the predicted fraction of infections $\hat{\mathcal{I}}_i[n+1], ..., \hat{\mathcal{I}}_i[n+n_\textrm{pred}]$ for all cities $i$. 

\begin{algorithm}
	\caption{\texttt{Epidemic outbreak prediction}}
	\begin{algorithmic}[1]
		\State \textbf{Input: } reported fraction of infections $\mathcal{I}_{\textrm{rep}, i}[1], ..., \mathcal{I}_{\textrm{rep}, i}[n]$ for all cities $i$; prediction time $n_\textrm{pred}$
		\State \textbf{Output: } predicted fraction of infections $\hat{\mathcal{I}}_i[n+1], ..., \hat{\mathcal{I}}_i[n+n_\textrm{pred}]$ for all cities $i$ 	
    \phase{Data preprocessing}
		\State $\mathcal{I}_{\textrm{rep}, 1}[17]\gets (\mathcal{I}_{\textrm{rep}, 1}[16] + \mathcal{I}_{\textrm{rep}, 1}[18])/2$			
		\State $\mathcal{I}_i[1], ..., \mathcal{I}_i[n] \gets$ \texttt{smoothdata}$(\mathcal{I}_{\textrm{rep}, i}[1], ..., \mathcal{I}_{\textrm{rep}, i}[n])$ for all $i=1, ..., N$
\State $\mathcal{I}[k] \gets (\mathcal{I}_1[k], ..., \mathcal{I}_N[k])^T$ for all $k=1, ..., n$	
\phase{Network inference}
		\For {$i=1, ..., N$}
\State $\mathcal{R}_i[1]\gets 0$	
		\For {$\delta_i \in \Omega$}
		\State $\mathcal{R}_i[k] \gets \mathcal{R}_i[k-1] + \delta_i \mathcal{I}_i[k-1]$ for all $k=2, ..., n$		
		\State $\mathcal{S}_i[k] \gets 1- \mathcal{I}_i[k] - \mathcal{R}_i[k]$ for all $k=1, ..., n$
\State $v_i[k]\gets (\mathcal{S}_i[k], \mathcal{I}_i[k],\mathcal{R}_i[k])^T$ for all $k=1, ..., n$
\State $(\beta_{i1}\left( \delta_i \right), ..., \beta_{iN}\left( \delta_i \right), \operatorname{MSE}\left( \delta_i \right) )\gets$ \texttt{Network inference}$\left( \delta_i, v_i[1], ..., v_i[n], \mathcal{I}[1], ..., \mathcal{I}[n]\right)$		
		\EndFor	
   \State $\hat{\delta}_i \gets \underset{\delta_i \in \Omega}{\operatorname{arg min}}\operatorname{MSE}\left(\delta_i\right)$
	\State $(\hat{\beta}_{i1}, ..., \hat{\beta}_{iN}) \gets \beta_{i1}( \hat{\delta}_i), ..., \beta_{iN} (\hat{\delta}_i)$ 
		\EndFor		
\phase{Iterating SIR model}		
\For {$i=1, ..., N$}
		\State $\hat{\mathcal{I}}_i[n]\gets\mathcal{I}_i[n]$
\State $\hat{\mathcal{R}}_i[1]\gets 0$	
		\State $\hat{\mathcal{R}}_i[k] \gets \hat{\mathcal{R}}_i[k-1] + \hat{\delta}_i \mathcal{I}_i[k-1]$ for all $k=2, ..., n$		
	\EndFor
		\For {$k=n+1, ..., n+n_\textrm{pred}$}		
		\For {$i=1, ..., N$}
\State $\hat{\mathcal{I}}_i[ k ] \gets (1 - \hat{\delta}_i) \hat{\mathcal{I}}_i[k-1] + (1 - \hat{\mathcal{I}}_i[k-1] - \hat{\mathcal{R}}_i[k-1]) \sum^N_{j=1} \hat{\beta}_{ij} \hat{\mathcal{I}}_j[k-1]$
	\State $\hat{\mathcal{R}}_i[ k ] \gets \hat{\mathcal{R}}_i[k-1] +  \hat{\delta}_i \hat{\mathcal{I}}_i[k-1]$	
	\EndFor
		\EndFor	
	\end{algorithmic}
	\label{algorithm:prediction}
\end{algorithm}

To determine the regularisation parameter $\rho_i$ in the LASSO (\ref{lasso}), we consider 100 candidate values specified by the set $\Theta_i=\{\rho_{\textrm{min}, i}, ..., \rho_{\textrm{max}, i}\}$. In line~4 of Algorithm~\ref{algorithm:reconstruction}, the maximum value is set to $\rho_{\textrm{max}, i}= 2 \lVert F^T_i V_i \rVert_\infty$. If $\rho_i>\rho_{\textrm{max}, i}$, then \cite{kim2007interior} the solution to the LASSO (\ref{lasso}) is $\beta_{ij}=0$ for all cities~$j$. For every value of the regularisation parameter $\rho_i \in \Theta_i$, we compute the mean squared error $\operatorname{MSE}\left( \delta_i, \rho_i \right)$ by 3-fold cross--validation \cite{hastie2015statistical}. For every fold, the rows of the matrix $F_i$ and the vector $V_i$ are divided into a training set $F_{i, \textrm{train}}$, $V_{i, \textrm{train}}$ and a validation set $F_{i, \textrm{val}}$, $V_{i, \textrm{val}}$. We compute the solution $\beta_{i1}$, ..., $\beta_{iN}$ to the LASSO (\ref{lasso}) on the training set of every fold $F_{i, \textrm{train}}$, $V_{i, \textrm{train}}$. The mean squared error $\operatorname{MSE}\left( \delta_i, \rho_i \right)$ then equals
\begin{align}\nonumber
\left\lVert V_{i, \textrm{val}} - F_{i, \textrm{val}} \begin{pmatrix}
\beta_{i1} \\
\vdots \\
 \beta_{iN}
\end{pmatrix}\right\rVert^2_2,
\end{align}
averaged over all folds. Finally, we set the regularisation parameter $\rho_i$ to the minimiser of $\operatorname{MSE}\left( \delta_i, \rho_i \right)$. The final estimate $\beta_{i1}( \delta_i ), ..., \beta_{iN} ( \delta_i )$ for the infection probabilities is obtained by solving the LASSO~(\ref{lasso}) on the whole matrix $F_i$ and vector $V_i$. To solve the LASSO (\ref{lasso}) numerically, we make use of the Matlab command \texttt{quadprog}.

\begin{algorithm}
	\caption{\texttt{Network inference}}
	\begin{algorithmic}[1]
		\State \textbf{Input: }  curing probability $\delta_i$; viral state $v_i[k]$ for $k=1, ..., n$; infection state vector $\mathcal{I}[k]$ for $k=1, ..., n$
		\State \textbf{Output: } infection probability estimates $\beta_{i1}( \delta_i ), ..., \beta_{iN}( \delta_i )$; mean squared error $\operatorname{MSE}( \delta_i )$	
		\State Compute $V_i$ and $F_i$ by (\ref{V_i}) and (\ref{F_i})		
		\State $\rho_{\textrm{max}, i}\gets 2 \lVert F^T_i V_i \rVert_\infty$
		\State $\rho_{\textrm{min}, i}\gets  ~10^{-4}\rho_{\textrm{max}, i}$
		\State $\Theta_i \gets$ 100 logarithmically equidistant values from $\rho_{\textrm{min}, i}$ to $\rho_{\textrm{max}, i}$				
		\For {$\rho_i \in \Theta_i$}			
		\State estimate $\operatorname{MSE}( \delta_i, \rho_i )$ by 3-fold cross--validation on $F_i, V_i$ and solving (\ref{lasso}) on the respective training set
		\EndFor
		\State $\rho_{\textrm{opt}, i} \gets \underset{\rho_i \in \Theta_i}{\operatorname{arg min}}\operatorname{MSE}\left(\delta_i, \rho_i\right)$
	\State $(\beta_{i1}( \delta_i ), ..., \beta_{iN} ( \delta_i ))\gets$ the solution to (\ref{lasso}) on the whole data set $F_i, V_i$ for $\rho_i = \rho_{\textrm{opt}, i}$
	\State $\operatorname{MSE}( \delta_i )\gets \operatorname{MSE}(\delta_i, \rho_{\textrm{opt}, i})$
	\end{algorithmic}
	\label{algorithm:reconstruction}
\end{algorithm}

\section{Logistic Regression} \label{appendix:logistic_regression}
A logistic curve is given by the following equation
\begin{equation}
	y(t) = \frac{y_\infty}{1+e^{-K(t-t_0)}}. \label{eq:logistic equation}
\end{equation}
In our formulation, $y(t)$ is the time-dependent fraction of infectious individuals, $t$ is the time in days, where January 21 serves as initial condition ($t=0$), $y_\infty$ is the fraction of infected individuals when time approaches infinity, $K$ is the logistic growth rate and $t_0$ indicates the inflection point of the logistic equation. For each city in Hubei, we have applied the Matlab command \texttt{lsqcurvefit} to fit the reported cumulative fraction 
\begin{align}\nonumber
\mathcal{I}_{\textrm{rep},\textrm{cs}, i} [k] = \sum^k_{\tau=1} \mathcal{I}_{\textrm{rep}, i} [k]
\end{align}
of infected individual to equation (\ref{eq:logistic equation}). 

\end{document}